\documentclass{article}
\input{tcilatex}

\begin{document}

\begin{center}
FRACTAL POWER LAW IN LITERARY ENGLISH$^{\S }$

\bigskip

L. L. Gon\c{c}alves$^{^{\dag }}$

Departamento de Fisica Geral,

Universidade de S\~{a}o Paulo,

C.P. 66318, 05315-970, S\~{a}o Paulo, SP, Brazil

and

L. B. Gon\c{c}alves$^{\dag \dag }$

Departamento de Letras Modernas

Universidade de S\~{a}o Paulo

C.P. 2530, 01060-970, S\~{a}o Paulo, SP, Brazil

\bigskip

Abstract
\end{center}

We present in this paper a numerical investigation of literary texts by
various well-known English writers, covering the first half of the twentieth
century, based upon the results obtained through corpus analysis of the
texts. A fractal power law is obtained for the lexical\ wealth defined as
the ratio between the number of different words and the total number of
words of a given text. By considering as a signature of each author the
exponent and the amplitude of the power law, and the standard deviation of
the lexical wealth, it is possible to discriminate works of different genres
and writers and show that each writer has a very distinct signature, either
considered among other literary writers or compared with writers of
non-literary texts. It is also shown that, for a given author, the signature
is able to discriminate between short stories and novels.

$\bigskip $


\ \ \ \ \ \ \ \ \ \ \ \ \ \ \ fractal power law, literary English.


$\bigskip $

$^{\dagger }$Corresponding author.

E-mail address: lindberg@fisica.ufc.br

On sabbatical leave from:

Departamento de Fisica

Universidade Federal do Cear\'{a},

Campus do Pici, C.P. 6030,

60451-970 Fortaleza, Cear\'{a}, Brazil

$\bigskip $

$^{\dag \dag }$On a post-graduate leave from:

Departamento de Letras Estrangeiras

Universidade Federal do Cear\'{a}

Av. da Universidade, 2683

60020-181, Fortaleza, Cear\'{a}, Brazil

\bigskip

$^{\S }$Work partially financed by the Brazilian agencies CNPq, Finep and
Capes.

\pagebreak

\textbf{1. INTRODUCTION}

The power law distribution of events introduced by the Italian economist
Pareto [1], in the context of the wealth of nations and individuals, and
restated by Zipf [2] concerning linguistics is perhaps the most ubiquitous
law in nature. This power law statistics, which is a characteristic of
fractal behaviour, is present in many different areas ranging from physics
[3] and biology [4,5] to natural hazards [6,7,8], and from musical creative
context [9] to economics [10]. Several applications in linguistics have
already been done, but the literary aspects which have been analysed until
now are all related to word formation within the literary texts [12,13]. To
the best of our knowledge, no attempt has been made to verify the
possibility of existence of a Zipf's law type within the literary context.
By using the software WordSmith, from Oxford University Press, with
electronic texts one is able to obtain the types/tokens ratio, types being
the number of different words in a text and tokens, the total number of
words in this text. Considering the fact that the types/tokens ratio will
necessarily decrease as the tokens increase, we were led to think of a power
law distribution, and therefore of relating this behaviour to Zipf's laws.

Although the use of a rich vocabulary is not the sole indication of
creativity in a writer, lexical wealth is certainly one characteristic to be
expected in a literary writer of the stature of J. Joyce, D. H. Lawrence, V.
Woolf., K. Mansfield and others. As a hypothesis, one would expect to find
in literary texts a high rate of type/token if compared with non-literary
texts, which means that the two genres should follow different Zipf's law.

A second hypothesis would be that, for the same literary author, the rate
type/token should be higher in the case of short stories, which belong to a
compact genre, rather than with novels, which are more prolix, and therefore
less creative in lexical terms. Then, we should be able to identify in
quantitative terms these differences.

Therefore, besides verifying a Zipf's law type of behaviour in different
texts, two important questions to ask are whether a `signature' may be
attributed to each author, by means of the exponent of the law, and if this
signature will change when the literary genre changes. Moreover, it would
also be desirable to associate different `signatures' to different authors,
even when they belong to the same literary period.

The purpose of this paper is to answer these questions, by analysing short
stories of famous writers, excerpts of novels, as well as non-literary
twentieth-century texts. To this aim, electronic corpora were compiled, with
15 short stories by J. Joyce, 12 short stories by D. H. Lawrence, 20 stories
by V. Woolf, 17 short stories by K. Mansfield and 24 short stories by
various authors. Also excerpts of five of Lawrence's novels and eight
excerpts from novels by different authors were compiled. We have also
considered 30 non-literary texts compiled from the newspaper The Guardian,
selected among various types of subjects.

In section 2 we present the works mentioned above and the results obtained
by using WordSmith tools. We also introduce Zipf's law in the literary
context. The results are presented in section 3 and the main conclusions are
summarized in the concluding remarks in section 4.

\bigskip\ 

\textbf{2. ZIPF'S LAW IN LITERATURE}

As described above, one important aspect of the literary work of a given
author is its lexical wealth, $k$, which is defined as the ratio between the
number of types $n$\ (number of different words) and the number of tokens $N$
(total number of words) of a given text. This quantity can be obtained from
the text analysis by using one of the lists, the S-list, given by the
WordList tool, provided by WordSmith. In the output we can obtain $n$,
identified as types, $N$, as tokens, and the ratio $K=100k(k=n/N$), which
corresponds to the lexical wealth in terms of percentage.

As pointed out in the Introduction, the lexical wealth represents the main
quantitative variable on which we will base our numerical analysis. By
assuming a power law behaviour for these quantities we can write

\bigskip 
\begin{equation}
N=Ak^{-\phi },  \tag{1}
\end{equation}

\bigskip

\noindent where $A$ is a constant amplitude and $\phi $ an exponent which
should be characteristics of a given author. We can linearize Eq. 1 by
taking the decimal logarithm on both sides of the equation, and from this we
obtain

\begin{equation}
\log _{10}(N)=\log _{10}(A)-\phi \log _{10}(k).  \tag{2}
\end{equation}%
\bigskip

The previous equation is a straight line in the new variables $\log _{10}(N)$
and $\log _{10}(k),$ and the constants $\alpha $ [$\alpha \equiv \log
_{10}(A)]\ $and $\phi $ are determined by means of a linear regression of
the data.

The data for different texts and authors analysed in this paper are shown in
Tables 1-8. \ In Tables 1-4 we present the short stories data relative to
the authors J. Joyce, D. H. Lawrence, V. Woolf and K. Mansfield,
respectively. The short stories data for various writers are shown in Table
5, and the excerpts of novels data by D. H. Lawrence and by different
authors are shown in Tables 6 and 7, respectively. Finally, in Table 8, we
present the data for non-literary texts, all obtained from recent issues of
the newspaper The Guardian. It should be noted that the rather large values
of $K$, for non-literary texts shown in this table, when compared with the
literary ones points to an apparent high\ lexical wealth which is a
consequence of the relatively small values of $N$ for the articles
considered.

The texts and authors have been chosen in such a way that we could have very
specific literary genres and representatives writers. In particular in order
to characterize the novels, which constitute a very important genre, we have
considered excerpts from five of D. H. Lawrence's novels to verify whether
there was any change within the writings of a same author, depending on his
writing a short story or a novel. Eight different novelists were studied so
that we had a comparable number against which to judge the results with
Lawrence's novels. A comparison with non-literary was essential in the sense
that it would be a validation of our study and, therefore journalistic texts
were considered.

\begin{center}
\bigskip
\end{center}

\textbf{3. RESULTS AND DISCUSSION}

The results for the linear regression of the data corresponding to short
stories by J. Joyce , D. H. Lawrence, V. Woolf and K. Mansfield are shown in
Figs. 1-4 respectively, and in Fig. 5 are shown the results for short
stories by different authors. In Figs. 6 and 7 are presented the results for
the novels of D. H. Lawrence and novels of various authors respectively.
Finally in Fig.8 we present the data for non-literary texts.

The results show clearly that a power law is satisfied in all cases with
some acceptable fluctuation in the parameters $\alpha (\alpha \equiv \log
A)\ $and $\phi .$ In literary terms, we may consider $A$ and $\phi $ as
signatures of the authors. For example, we can observe that in the limit $%
k\rightarrow 1$ we have $N=A=n$, which gives, in theoretical terms, the
number of words an author may write before he repeats a word. The parameter $%
\phi $ will determine how rapidly, or slowly, repeated \ words are
introduced in the discourse, which is typical for each writer.

In the case of the short stories by different authors, as expected, the
dispersion of the data from the power law is more pronounced (Fig. 5), as we
were dealing with different signatures to the same work. The least
dispersions are presented by Lawrence and Mansfield's short stories, as can
be seen in Figs. 2 and 4, respectively.

Identical behaviour for different authors is also observed when the novels
are considered, as shown in Figs. 6 and 7, and the largest dispersion is
present in the non-literary texts shown in Fig. 8. This is an indication
that a measure of this dispersion, $\chi =100\eta $, where $\eta $ is the
standard deviation of the linear fitting, can be used as a parameter to
discriminate the authors and genres of literary works.

In the literary perspective, $\chi $ will indicate the consistency of a
writer, for any value of $k$, and it is remarkable in this aspect the result
for Lawrence's novels. It is worthwhile to consider that, when dealing with
short stories, where $N$ is usually small, the dispersion interferes more
substantially in the calculation of $\chi .$ In the newspaper texts,
although the context imposes a series of norms in writing, we can find a
high $\chi ,$ which identifies these texts as non-literary.

We can also define an important additional parameter, related to the
uniformity of the lexical wealth. This is the standard deviation of the $K$,
which we define as $\sigma .$ Certainly it constitutes a typical personal
parameter, since it is directly related to the lexical wealth of the author,
and it has been recently introduced in the context of the characterization
of climate of different regions in the United States from the analysis of
maximum daily temperature time series[14]. In literary terms, $\sigma $ will
have to do with the preservation of lexical wealth, and with different sizes
of the analized texts. Thus, the novels, either by D. H. Lawrence or by
different authors, will have lower $\sigma ^{\prime }s$ also due to the fact
that the excerpts had similar lengths.

The values of the indices for various authors and genres are summarized in
Table 9,\textbf{\ }and they will be used as their signatures. The
identification of the various authors and genres through fractal indices
follows a procedure which has been recently used in the description of
different complex systems. Examples of such a procedure can be found in our
recent paper on the characterization of acoustic emission signals [15], and
in various references quoted therein.

From the analysis of these values we can conclude that the various genres
tend to occupy separate regions of the four-dimensional space generated by
these parameters, as in the case of ref. [15]. This can be seen more clearly
in the projection of the points in different planes shown in Figs. 9-11, and
it should be noted that these projections do not represent functions of the
shown variables. Although this separation is not very well defined in the
diagram shown in Fig. 9 (a), where the novels are not completely segregated,
the separation of genres are very well marked in all the other diagrams
shown in Figs. 9(b), 10(a,b) and 11(a,b). This remarkable result can also be
seen in the three-dimensional plots, shown in Figs. 12-15, which correspond
to projections of the points on the three dimensional subspaces. And it
shows that, although various writers can be differentiated by their
signatures in the way shown above, the genres of their writing can be
identified as clearly, as it was observed for short stories, novels and
newspaper articles.

Although it was not the objective of our study, we have verified that the
charateristic parameters for power law, which relates the total number of
letters of given text with the number of words, is dependent on the size of
the text. The exponent of the power law, which is equal to one [16], can
only be obtained correctly when we consider statistically representative
samples which is consistent with the results of P\"{o}schel et al.[17]

In order to verify the presence of similar effect on our characteristic
parameters, we have analysed excerpts of Lawrence's novels with different
sizes. Differently from the previous power law, which is defined in the
context of linguistics, no significant variation in the parameters was
observed. This is an indication that the usual statistical approach used in
linguistic should not be applied to study of literary texts.

Finally, we can conclude from the results presented in the Figs. 9-15 that,
besides identifying the genres, the parameters $\chi ,$ $\sigma $, $\alpha $
and $\phi $, are also characteristic of each author. In particular it should
be noted that even within a literary genre, short stories or novels, these
indices discriminate the work of different authors as well as a collection
of works by many authors. Therefore, they constitute the literary signature
of the genres, authors and group of authors.

\bigskip

\textbf{4. CONCLUDING REMARKS}

We have shown, from the corpus analysis of the literary production of
English authors and non-literary texts, that a power fractal law can be
associated with the lexical wealth of the authors. The parameters defined in
the power law, the standard deviation of the power law fitting and the
standard deviation of the lexical wealth can be used to characterize each
writer and genre and to discriminate a literary from a non-literary text.
This discrimination among authors, different literary texts, namely, short
stories and novel, and non-literary articles is present in all the
two-dimensional diagrams built with the parameters $\phi ,$ $\alpha $, $\chi 
$ and $\sigma ,$ as has been shown in Figs. 9-11. The three dimensional plot
constructed with these parameters are shown in Figs.12-15, and they clearly
present the separation among the different authors and styles.

We believe, from the results presented in this work, that the statistical
analysis introduced by Corpus Linguistic can increase objectivity in
literary studies, opening a wide field for further research.

In order to give further support to this conclusion, following Havlin [18],
we introduce a distance from Zipf plots for a given set of data of the
authors and genres as

\begin{equation}
d_{AB}=\sqrt{\frac{1}{N_{A}N_{B}}\dsum\limits_{i,j}{\LARGE (}\overline{K}%
_{i}^{A}-\overline{K}_{j}^{B}{\LARGE )}^{2}},  \tag{3}
\end{equation}%
where A and B identify the authors and genres,$\ i$ and $j$\ the different
works in a given genre, \ $N_{A(B)}$ is the total number of works considered
for a given author and genre, and $\overline{K}_{\beta }^{\alpha }$ is given
by

\begin{equation}
\mathbf{\ }\overline{K}_{\beta }^{\alpha }=\mathbf{\ }K_{\beta }^{\alpha }-%
\mathbf{\ }\widetilde{K}_{\beta }^{\alpha },  \tag{4}
\end{equation}%
where $\widetilde{K}_{\beta }^{\alpha }$ is obtained from the linear
regression of Eq. (2). Eq.(3) can also be applied when the authors A and B
are identical, and, in this case, the distance $d_{\alpha \alpha }$\
measures the deviation of the data from a Zipf plot.

The results obtained are presented in Table 10 which, by construction,
corresponds to a symmetric matrix. As expected, the shortest distances
correspond to the ones related to a same author, and, for a given author,
there are different distances for different authors and genres. These
results are similar to the ones obtained by Vilensky [19] in the context of
Linguistics and gives support to the proposed discrimination within our
fractal analysis.

\bigskip

\textbf{ACKNOWLEDGEMENTS}

The authors would like to thank Dr. A. P. Vieira and Dr. J. P. de Lima for
relevant remarks and a critical reading of the manuscript. We would also
like to thank the anonymous referee for making us aware of the works by
Havlin [18] and Vilensky [19] on the definition and use of the concept of
distance among Zipf plots.

\bigskip

\textbf{REFERENCES}

1. V. Pareto, \textit{Cours d'\'{e}conomie politique profess\'{e} \`{a}
l'Universit\'{e} de Lausanne},

\ \ \ (Rouge, 1897)

2. G. K. Zipf, \textit{Selective Studies and the Principle of Relative
Frequency in }

\ \ \ \textit{Language} (Harvard University Press, 1932)

\ \ \ G. K. Zipf, \textit{Psycho-Biology of Languages} (Houghton-Mifflin,
1935;

\ \ \ MIT Press, 1965).

\ \ \ G. K. Zipf, \textit{Human Behavior and the Principle of Least Effort}

\ \ \textit{\ }(Addison-Wesley, 1949).

3.D. Sornette, \textit{Critical phenomena in natural sciences:chaos,
fractals, }

\ \ \ s\textit{elf-organization and disorder: concepts and tools}, 2nd \
edn. (Springer,

\ \ 2004)

4. V. V. Solovyev, Biosystems \textbf{30}, 137 (1993)

5. B. Suki, A. L. Barab\'{a}si, Z. Hantos, F. Pet\'{a}k, and H. E. Stanley, 
\textit{Nature}

\ \ \ \textbf{368}, 615 (1994)

6 S. Hergarten, \textit{Natural Hazards and Earth System Sciences} \textbf{4}%
,

\ \ \ 309( 2004)

7 B.D. Malamud, D. L. Turcotte, F. Guzetti and P. Reichenbach, \textit{Earth 
}

\ \ \textit{Surface Processes and Landforms }\textbf{29}, 309 (2004)

8 D. L. Turcotte, \textit{Fractals and Chaos in Geology and Geophysics,} 2nd
edn

\ \ (Cambridge University Press, 1997)

9. D. H. Zanette, cs.CL/0406015

10 R. N. Mantegna and H. E Stanley, \textit{An Introduction to Econophysics: 
}

\ \ \ \ \textit{Correlations and Complexity in Finance }(Cambridge
University Press,

\ \ \ \ 1999)

11. A. Gelbukh and G. Sidorov, \textit{Lecture Notes in Computer Science}%
\textbf{\ 2004},

\ \ \ \ \ 332 (2001)

12. W. Ebeling and T. P\"{o}schel, \textit{Europhys. Lett}. \textbf{26}, 241
(1994)

13. A. Eftekhari, \textit{cs.CL/0408041}

14. M.L. Kurnaz, \textit{J. Stat. Mech.: Theor. Exp}. P07009 (2004)

15. F.E. Silva, L.L. Gon\c{c}alves, D.B.B. Ferreira and J.M.A.\ Rebello, 
\textit{Chaos,}

\ \ \ \ \textit{\ Solitons \& Fractals }\textbf{26}, 481 (2005).

16. R. Perling, \textit{Phys. Rev. B} \textbf{54}, 220 (1996).

17. T. P\"{o}schel, W. Ebeling, C. Fr\"{o}mmel and R. Ramirez, Eur. Phys. J.

\ \ \ \ \ E \textbf{12}, 531 (2003).

18. S. Havlin,\textit{\ Physica A} \textbf{215}, 148 (1995).

19. B. Vilensky, \textit{Physica A} \textbf{231}, 705 (1995).

\bigskip

\textbf{FIGURE CAPTIONS}

\bigskip

Fig. 1 - Linear fit of $log_{10}(N)$ as a function of $log_{10}(k)$ for
short stories

\ \ \ \ \ \ \ \ \ \ by J. Joyce.

Fig. 2 - Linear fit of $log_{10}(N)$ as a function of $log_{10}(k)$ for
short stories

\ \ \ \ \ \ \ \ \ \ by D. H. Lawence.

Fig. 3 - Linear fit of $log_{10}(N)$ as a function of$log_{10}(k)$ for short
stories

\ \ \ \ \ \ \ \ \ \ by V. Woolf.

Fig. 4 - Linear fit of $log_{10}(N)$ as a function of $log_{10}(k)$ for
short stories

\ \ \ \ \ \ \ \ \ \ by K. Mansfield.

Fig. 5 - Linear fit of $log_{10}(N)$ as a function of $log_{10}(k)$ for
short stories

\ \ \ \ \ \ \ \ \ \ by various authors.

Fig. 6 - Linear fit of $log_{10}(N)$as a function of $log_{10}(k)$ for
excerpts of

\ \ \ \ \ \ \ \ \ \ novels by D. H. Lawrence.

Fig. 7 - Linear fit of $log_{10}(N)$ as a function of $log_{10}(k)$ for
excerpts of

\ \ \ \ \ \ \ \ \ \ novels by various authors.

Fig. 8 - Linear fit of $log_{10}(N)$as a function of $log_{10}(k)$ for
excerpts of

\ \ \ \ \ \ \ \ \ \ non-literary texts.

Fig. 9 - (a) Percentage standard deviation of the linear fitting, $\chi ,$
and

\ \ \ \ \ \ \ \ \ \ \ \ (b) logarithm of the amplitude of the\ power law, $%
\alpha ,$ as functions

\ \ \ \ \ \ \ \ \ \ \ \ of the power law exponent, $\phi ,$ for different
authors,genres and

\ \ \ \ \ \ \ \ \ \ \ \ non-literary texts.

Fig. 10 - Standard deviation of $K$, $\sigma ,$ (a) as a function of the
power law

\ \ \ \ \ \ \ \ \ \ \ \ exponent, $\phi ,$ and (b) as a function of the the
percentage standard

\ \ \ \ \ \ \ \ \ \ \ \ deviation of the linear fitting, $\chi ,$ for
different authors, genres\ 

\ \ \ \ \ \ \ \ \ \ \ \ and non-literary texts.

Fig 11 - Logarithm of the amplitude of the\ power law, $\alpha ,$ (a) as a
function

\ \ \ \ \ \ \ \ \ \ \ \ of the standard deviation of $K,$ $\sigma ,$ and (b)
as a function of the

\ \ \ \ \ \ \ \ \ \ \ percentage standard deviation of the linear fitting, $%
\chi ,$ for different

\ \ \ \ \ \ \ \ \ \ \ authors, genres and non-literary texts.

Fig. 12 - \ Standard deviation of $K,$ $\sigma $ , as a function of the
power law

\ \ \ \ \ \ \ \ \ \ \ \ \ exponent, $\phi ,$ and of the logarithm of the
amplitude of the\ power

\ \ \ \ \ \ \ \ \ \ \ law, $\alpha ,$ for different authors, genres and
non-literary texts.

Fig. 13 - The standard deviation of $K,$ $\sigma $ , as a function of the
power law

\ \ \ \ \ \ \ \ \ \ \ \ \ exponent, $\phi ,$ and of the percentage\ standard
deviation\ of the

\ \ \ \ \ \ \ \ \ \ \ \ linear fitting $\chi ,$ for different authors,
genres and\ non-literary texts.

Fig. 14 - The logarithm of the amplitude of the\ power law, $\alpha ,$ as a
function

\ \ \ \ \ \ \ \ \ \ \ \ \ of the percentage standard deviation of the linear
fitting, $\chi ,$ and of

\ \ \ \ \ \ \ \ \ \ \ \ the power law exponent, $\phi ,$\ for different
authors,\ genres and

\ \ \ \ \ \ \ \ \ \ \ \ non-literary texts.

Fig. 15 - The logarithm of the amplitude of the\ power law, $\alpha ,$\ as a

\ \ \ \ \ \ \ \ \ \ \ \ \ \ function of the percentage\ standard deviation
of the linear

\ \ \ \ \ \ \ \ \ \ \ \ \ \ fitting $\chi $ and of the standard\ deviation\
of $K,$ $\ \sigma ,$ for \ different

\ \ \ \ \ \ \ \ \ \ \ \ \ authors, genres and non-literary texts.

\bigskip \pagebreak

\bigskip \textbf{TABLES}

\begin{tabular}{|c|c|c|c|}
\hline
Titles & $N$ & $n$ & $K$ \\ \hline
After the race & 2255 & 853 & 37.83 \\ \hline
A little cloud & 5063 & 1361 & 26.88 \\ \hline
A mother & 4605 & 1177 & 25.56 \\ \hline
An encounter & 3281 & 976 & 29.75 \\ \hline
A painful case & 3672 & 1231 & 33.52 \\ \hline
Araby & 2367 & 822 & 34.73 \\ \hline
Clay & 2686 & 719 & 26.77 \\ \hline
Counterparts & 4201 & 1092 & 25.99 \\ \hline
Eveline & 1842 & 625 & 33.93 \\ \hline
Grace & 7745 & 1766 & 22.80 \\ \hline
Ivy day in the committee room & 5487 & 1210 & 22.05 \\ \hline
The boarding house & 2850 & 933 & 32.74 \\ \hline
The dead & 16006 & 2702 & 16.88 \\ \hline
The sisters & 3169 & 888 & 28.02 \\ \hline
Two gallants & 3984 & 1134 & 28.46 \\ \hline
\end{tabular}

Table 1. Short stories by J. Joyce. $N$ is the number of\ tokens,

\ \ \ \ \ \ \ \ \ \ \ \ \ $\ n$ is the number of types and $K=100n/N$.

\bigskip

\begin{center}
\bigskip
\end{center}

\begin{tabular}{|c|c|c|c|}
\hline
Titles & $N$ & $n$ & $K$ \\ \hline
A fragment of stained glass & 4044 & 1046 & 25.87 \\ \hline
A sick collier & 2505 & 792 & 31.62 \\ \hline
Daughters of the vicar & 19902 & 3012 & 15.13 \\ \hline
Goose fair & 3990 & 1180 & 29.57 \\ \hline
Older of chrysanthemums & 7806 & 1671 & 21.41 \\ \hline
Second best & 3095 & 997 & 32.21 \\ \hline
The christening & 3604 & 1099 & 30.49 \\ \hline
The Prussian officer & 9163 & 1793 & 19.57 \\ \hline
The shades of spring & 5477 & 1397 & 25.51 \\ \hline
The shadow in the rose garden & 4912 & 1175 & 23.92 \\ \hline
The thorn in the flesh & 7303 & 1708 & 23.39 \\ \hline
The white stocking & 8379 & 1684 & 20.10 \\ \hline
\end{tabular}

Table 2. Short Stories by D. H. Lawrence. $N$ is the number

\ \ \ \ \ \ \ \ \ \ \ \ \ of\ tokens, $n$ is the number of types and $%
K=100n/N$.

\bigskip

\bigskip

\bigskip

\begin{tabular}{|c|c|c|c|}
\hline
Titles & $N$ & $n$ & $K$ \\ \hline
A summing up & 1340 & 554 & 41.34 \\ \hline
An unwritten novel & 4667 & 1393 & 29.85 \\ \hline
A woman's college from outside & 1321 & 568 & 43.00 \\ \hline
In the orchard & 1008 & 417 & 41.37 \\ \hline
Kew gardens & 2656 & 902 & 33.96 \\ \hline
Lappin and Lappinova & 3259 & 997 & 30.59 \\ \hline
Moments of being & 2728 & 837 & 30.68 \\ \hline
Mrs. Dalloway in Bond Street & 3139 & 986 & 31.41 \\ \hline
Solid objects & 2415 & 859 & 35.57 \\ \hline
Society & 5472 & 1414 & 25.84 \\ \hline
The duchess and the jeweller & 2505 & 823 & 32.85 \\ \hline
Together and apart & 2288 & 764 & 33.39 \\ \hline
The legacy & 3062 & 797 & 29.03 \\ \hline
The lady in the looking glass & 2231 & 746 & 33.44 \\ \hline
The man who loved his kind & 2385 & 738 & 30.94 \\ \hline
The mark on the wall & 3178 & 1071 & 33.70 \\ \hline
The new dress & 3276 & 949 & 28.97 \\ \hline
The searchlight & 1617 & 486 & 30.06 \\ \hline
The shooting party & 2954 & 896 & 30.33 \\ \hline
The string quartet & 1511 & 726 & 48.05 \\ \hline
\end{tabular}

Table 3. Short Stories by V. Woolf.\ $N$ is the number \ of\ tokens,

\ \ \ \ \ \ \ \ \ \ \ \ \ $n$is the number of types and $K=100n/N$.

\bigskip

\begin{tabular}{|c|c|c|c|}
\hline
Titles & $N$ & $n$ & $K$ \\ \hline
An ideal family & 2524 & 786 & 31.14 \\ \hline
At the bay & 13746 & 2438 & 17.74 \\ \hline
Bliss & 4898 & 1170 & 23.89 \\ \hline
Bank holiday & 1264 & 582 & 46.06 \\ \hline
Her first ball & 2634 & 801 & 30.41 \\ \hline
Life of Ma Parker & 2643 & 778 & 29.44 \\ \hline
Marriage \`{a} la mode & 3925 & 1075 & 27.39 \\ \hline
Miss Brill & 2020 & 672 & 33.27 \\ \hline
Mr. and Mrs. Dove & 3537 & 892 & 25.22 \\ \hline
Prelude & 17123 & 2678 & 15.64 \\ \hline
The daughters of the late colonel & 7253 & 1404 & 19.36 \\ \hline
The garden party & 5567 & 1256 & 22.56 \\ \hline
The lady's maid & 2239 & 560 & 25.01 \\ \hline
The singing lesson & 2137 & 681 & 31.87 \\ \hline
The stranger & 4694 & 1064 & 22.67 \\ \hline
The voyage & 3224 & 909 & 28.19 \\ \hline
The young girl & 2278 & 712 & 31.26 \\ \hline
\end{tabular}

Table 4. Short Stories by K. Mansfield.\ $N$ is the number of\ tokens,

\ \ \ \ \ \ \ \ \ \ \ \ $n$ is the number of types and $K=100n/N$.

\begin{quote}
\bigskip
\end{quote}

\bigskip 
\begin{tabular}{|c|c|c|c|c|}
\hline
Titles & Authors & $N$ & $n$ & $K$ \\ \hline
A girl in it & R. Kenney & 4248 & 1296 & 30.51 \\ \hline
A hedonist & J. Galsworthy & 2760 & 1018 & 36.88 \\ \hline
Broadsheet ballad & A. E. Coppard & 2824 & 738 & 26.13 \\ \hline
Empty arms & R. Pertwee & 5083 & 1367 & 26.89 \\ \hline
Genius & E. Mordaunt & 8767 & 1865 & 21.27 \\ \hline
Lena Wrace & M. Sinclair & 4931 & 1073 & 21.76 \\ \hline
Major Willbraham & H. Walpole & 6425 & 1463 & 22.77 \\ \hline
Once a hero & H. Brighouse & 5946 & 1401 & 23.56 \\ \hline
Seaton's aunt & W. de la Mare & 12100 & 2516 & 20.79 \\ \hline
The backstairs of the mind & R. Langbridge & 1624 & 614 & 37.81 \\ \hline
The bat and Belfry Inn & A. Graham & 5527 & 1459 & 26.40 \\ \hline
The birth of a masterpiece & L. Malet & 4284 & 1545 & 36.06 \\ \hline
The christmas present & R. Crompton & 1556 & 533 & 34.25 \\ \hline
The devil to pay & M. Pemberton & 5092 & 1287 & 25.27 \\ \hline
The dice thrower & S. Southgate & 3138 & 971 & 30.94 \\ \hline
The lie & H. Horn & 863 & 365 & 42.29 \\ \hline
The looking-glass & J. D. Beresford & 8522 & 1742 & 20.44 \\ \hline
The olive & A. Blackwood & 4076 & 1264 & 31.01 \\ \hline
The pensioner & W. Caine & 1650 & 570 & 34.55 \\ \hline
The reaper & D. Easton & 1581 & 652 & 41.24 \\ \hline
The song & M. Edginton & 4688 & 1206 & 23.75 \\ \hline
The stranger woman & G. B. Stern & 5264 & 1608 & 46.36 \\ \hline
The woman who sat still & P. Trustcott & 2945 & 869 & 39.60 \\ \hline
Where was Wych Street & S. Aumonier & 5929 & 1617 & 27.27 \\ \hline
\end{tabular}

Table 5. Short Stories by various authors.\ $N$ is the number

\ \ \ \ \ \ \ \ \ \ \ \ \ of\ tokens, $n$ is the number of types and $%
K=100n/N$.

\bigskip

\bigskip

\begin{tabular}{|c|c|c|c|}
\hline
Titles & $N$ & $n$ & $K$ \\ \hline
Kangoroo & 18609 & 3131 & 16.83 \\ \hline
Lady Chatterley's lover & 32293 & 4154 & 12.86 \\ \hline
Sons and lovers & 33821 & 4395 & 12.99 \\ \hline
The rainbow & 29322 & 3982 & 13.58 \\ \hline
Women in love & 38964 & 4828 & 12.39 \\ \hline
\end{tabular}

Table 6. Excerpts of novels by D. H. Lawrence.\ $N$ is the number

\ \ \ \ \ \ \ \ \ \ \ \ \ of\ tokens, $n$ is the number of types and $%
K=100n/N$.

\bigskip

\bigskip

\begin{tabular}{|c|c|c|c|c|}
\hline
Titles & Authors & $N$ & $n$ & $K$ \\ \hline
Allan Quartermaine & H. R. Haggard & 15847 & 2960 & 18.68 \\ \hline
Jeremy and Hamlet & H. S. Walpole & 34007 & 189594 & 11.90 \\ \hline
Kim & R. Kipling & 24730 & 4041 & 16.34 \\ \hline
Men are like gods & H. G. Wells & 35760 & 5373 & 15.03 \\ \hline
Mrs. Dalloway & V. Woolf & 40054 & 5446 & 13.60 \\ \hline
The moon and sixpence & W. S. Maugham & 20334 & 3311 & 16.28 \\ \hline
The secret of Father Brown & G. K. Chesterton & 48610 & 5392 & 11.09 \\ 
\hline
The white monkey & J. Galsworthy & 46105 & 5576 & 12.09 \\ \hline
\end{tabular}

Table 7. Excerpts of novels by various authors. $N$ is the number

\ \ \ \ \ \ \ \ \ \ \ \ \ of\ tokens, $n$ is the number of types and $%
K=100n/N$.

\bigskip

\begin{tabular}{|c|c|c|c|c|}
\hline
Titles & Authors & $N$ & $n$ & $K$ \\ \hline
2001 Odyssey & R. Bray & 1651 & 727 & 44.03 \\ \hline
Arsenal boss? I could be & I. Ridley & 2253 & 648 & 28.76 \\ \hline
Blair claims a Tory crown & K. Ahmed & 302 & 185 & 61.26 \\ \hline
Briton among 20 killed in ambush & J. Steel & 449 & 237 & 52.78 \\ \hline
Bubbles bursts & J. Martinson et al. & 1143 & 504 & 44.09 \\ \hline
Child of a dream & M. Benn & 1605 & 630 & 39.25 \\ \hline
Coal makes Oslo king of the isles & P. Brown & 939 & 419 & 44.62 \\ \hline
Courts make Bush confront race issue & D. Campbell & 575 & 289 & 49.74 \\ 
\hline
Dodgy wigs, extravagant costumes... & J. Mortimer & 1374 & 592 & 43.09 \\ 
\hline
Ferrying around & P. Daoust & 1499 & 670 & 44.70 \\ \hline
Foyer to the future & J. Freedland & 1229 & 537 & 43.69 \\ \hline
French bons vivants take a battering & J. Henley & 811 & 426 & 52.53 \\ 
\hline
Full fifth round draw & D. Fifield & 292 & 175 & 59.93 \\ \hline
Get your kit out & N. Holford & 203 & 118 & 58.13 \\ \hline
Green light for Sorrel \pounds 35m & D. Atkinson & 570 & 277 & 48.60 \\ 
\hline
Hitchcock veteran finally makes his .... & M. Kennedy & 481 & 241 & 50.10 \\ 
\hline
India and Pakistan are ... & M. Tran & 778 & 419 & 53.86 \\ \hline
Love-hate relationship with Woodhead & R. Allison & 313 & 184 & 59.79 \\ 
\hline
Mandelson resigns & J. O`Farrel & 825 & 414 & 50.18 \\ \hline
Neighbours in need & D. Brown & 734 & 400 & 54.50 \\ \hline
New quake scare & O. Bowcott & 376 & 222 & 59.04 \\ \hline
Partying in the pink & N. McIntosh & 1490 & 596 & 40.00 \\ \hline
Prosperous old capital strains ... & P. Hetherington & 637 & 317 & 49.76 \\ 
\hline
Racing & T. Paley & 218 & 136 & 62.39 \\ \hline
Rostrum & J. Brennan & 637 & 277 & 43.49 \\ \hline
Smoke screen & C. Arnot & 1014 & 465 & 45.86 \\ \hline
The pursuit of leisure & L. Brooks & 1671 & 796 & 47.64 \\ \hline
Theft and murder in the Balkans & D. Brown & 726 & 399 & 54.96 \\ \hline
Thumbs up for Scott's Gladiator & D. Campbell et al. & 836 & 380 & 45.45 \\ 
\hline
Two-minute fight for BA2069 & J. Vasagar et al. & 760 & 328 & 43.16 \\ \hline
\end{tabular}

Table 8. Non-literary texts. $N$ is the number of\ tokens,

\ \ \ \ \ \ \ \ \ \ \ \ \ \ $n$ is the number of types and $K=100n/N$.

\bigskip

\begin{tabular}{|c|c|c|c|c|}
\hline
Author/Genre & $\chi $ & $\phi $ & $\sigma $ & $\alpha $ \\ \hline
D. H. Lawrence & 6.12 & 2.39 & 5.52 & 2.31 \\ \hline
K. Mansfield & 9.56 & 2.58 & 6.88 & 2.10 \\ \hline
J. Joyce & 10.18 & 2.47 & 5.17 & 2.24 \\ \hline
V. Woolf & 11.85 & 2.07 & 5.55 & 2.41 \\ \hline
Short stories & 14.22 & 2.54 & 6.30 & 2.20 \\ \hline
D. H. Lawrence (novels) & 1.99 & 2.29 & 1.60 & 2.49 \\ \hline
Novels & 8.13 & 1.97 & 2.48 & 2.82 \\ \hline
Texts & 15.36 & 3.26 & 7.51 & 1.84 \\ \hline
\end{tabular}

Table 9. Characteristic indices of the various

\ \ \ \ \ \ \ \ \ \ \ \ \ authors and genres.

\bigskip

\bigskip

\bigskip

\bigskip

\bigskip

\bigskip

\bigskip

\begin{tabular}{|c|c|c|c|c|c|c|c|c|}
\hline
Author/Genre & (1) & (2) & (3) & (4) & (5) & (6) & (7) & (8) \\ \hline
D. H. Lawrence (1) & \textbf{2.08} & 32.93 & 30.27 & 22.46 & 32.22 & 40.99 & 
35.42 & 42.58 \\ \hline
K. Mansfield (2) & 32.93 & \textbf{3.39} & 31.28 & 22.17 & 31.74 & 44.28 & 
39.22 & 36.98 \\ \hline
J. Joyce (3) & 30.27 & 31.28 & \textbf{3.77} & 23.20 & 33.84 & 42.34 & 38.06
& 39.73 \\ \hline
V. Woolf (4) & 22.46 & 22.17 & 23.20 & \textbf{16.18} & 39.48 & 60.07 & 52.71
& 39.43 \\ \hline
Short stories (5) & 32.22 & 31.74 & 33.84 & 39.48 & \textbf{5.63} & 43.07 & 
36.93 & 27.85 \\ \hline
D. H. Lawrence (novels) (6) & 40.99 & 44.28 & 42.34 & 60.07 & 43.07 & 
\textbf{0.28} & 18.67 & 46.12 \\ \hline
Novels (7) & 35.42 & 39.22 & 38.06 & 52.71 & 36.93 & 18.67 & \textbf{1.66} & 
53.45 \\ \hline
Texts (8) & 42.58 & 36.98 & 39.73 & 39.43 & 27.85 & 46.12 & 53.45 & \textbf{%
7.33} \\ \hline
\end{tabular}

Table 10. Zipf distances among different authors and genres.

\bigskip 

\end{document}